\newif\ifNotes\Notesfalse
\newif\ifNCH\NCHfalse

\documentclass[10pt, conference, compsocconf]{IEEEtran}
\usepackage{fancyhdr}
\usepackage[normalem]{ulem}
\usepackage[hyphens]{url}
\usepackage[sort,nocompress]{cite}
\usepackage[final]{microtype}
\usepackage[keeplastbox]{flushend}
\usepackage{multirow}    
\usepackage{pifont}      
\usepackage{color}       
\usepackage{xspace}      
\usepackage{graphicx}
\usepackage{subcaption}
\usepackage{rotating}
\usepackage{paralist}
\let\labelindent\relax
\usepackage{enumitem}
\usepackage{soul}
\usepackage[linesnumbered,ruled]{algorithm2e}
\makeatletter
\renewcommand{\@algocf@capt@plain}{above}
\makeatother 
\usepackage{algorithmicx}  
\usepackage{algpseudocode}

\usepackage{amsmath,amsthm} 
\algnewcommand{\LineComment}[1]{\State \(\triangleright\) #1}
\usepackage{tikz}
\usetikzlibrary{arrows}
\usepackage{xcolor}
\usepackage{booktabs}
\usepackage[bookmarks=true,breaklinks=true,letterpaper=true,colorlinks,linkcolor=black,citecolor=blue,urlcolor=black]{hyperref}

\ifNotes
  \newcommand{\authcomment}[3]{\leavevmode\unskip\space\textcolor{#3}{#1 says: #2}\xspace}
  
\else
  \newcommand{\authcomment}[3]{\leavevmode\unskip\relax}
  
\fi

\ifNCH
  \newcommand{\NCH}{{\raisebox{0pt}{\color{red}\Large $\blacksquare$}}\xspace}
\else
  \newcommand{\NCH}{\leavevmode\unskip\relax\xspace}
\fi

\newcommand{\mypara}[1]{\vspace{2pt}\noindent\textbf{{#1: }}}
\newcommand{\eat}[1]{}

\newcommand{\UPPERTELE}{{Implicit Hammer}\xspace}

\newcommand{\upperperi}{{explicit hammer}\xspace}
\newcommand{\uppertele}{{implicit hammer}\xspace}
\newcommand{\upperpt}{PThammer\xspace}

\newcommand{\yval}[1]{\authcomment{yval}{#1}{purple}}

\newcommand{\us}{$\mu$s}

\SetKwFor{RepTimes}{repeat}{times}{end}
\newcommand{\misses}{\mathit{misses}}
\newcommand{\pts}{\mathit{profile\_tlb\_set}}

\newcommand{\loopnum}{\mathit{count}}
\newcommand{\targetaddr}{\mathit{target\_addr}}
\newcommand{\page}{\mathit{page}}
\newcommand{\set}{\mathit{set}}

\newcommand{\datamarker}{\mathit{data\_marker}}
\newcommand{\buf}{\mathit{buf}}
\newcommand{\initset}{\mathit{init\_set}}
\newcommand{\threshold}{\mathit{threshold}}
\newcommand{\temptlbmiss}{\mathit{temp\_tlb\_miss}}

\newcommand{\pes}{\mathit{profile\_evict\_set}}
\newcommand{\target}{\mathit{target}}
\newcommand{\memoryline}{\mathit{memory\_line}}
\newcommand{\latency}{\mathit{latency}}
\newcommand{\latencyset}{\mathit{latency\_set}}
\newcommand{\evictionsets}{\mathit{eviction\_sets}}
\newcommand{\pageoffset}{\mathit{page\_offset}}
\newcommand{\pteoffset}{\mathit{l1pte\_offset}}
\newcommand{\maxlatency}{\mathit{max\_latency}}
\newcommand{\maxset}{\mathit{max\_set}}

\usepackage{soul}
\usepackage[bordercolor=white,backgroundcolor=gray!30,linecolor=black,colorinlistoftodos]{todonotes}
\newcommand{\new}[1]{{\color{black}#1}\xspace}
\newcommand{\newpar}[1]{{\color{black}#1}\xspace}

\begin{document}
%
\title{PThammer: Cross-User-Kernel-Boundary Rowhammer through Implicit Accesses}


%
\author{\IEEEauthorblockN{Zhi Zhang\IEEEauthorrefmark{1}\IEEEauthorrefmark{2}\IEEEauthorrefmark{3},
Yueqiang Cheng\IEEEauthorrefmark{1}\IEEEauthorrefmark{4},
Dongxi Liu\IEEEauthorrefmark{3}, 
Surya Nepal\IEEEauthorrefmark{3},
Zhi Wang\IEEEauthorrefmark{5}, and
Yuval Yarom\IEEEauthorrefmark{3}\IEEEauthorrefmark{6}}
\IEEEauthorblockA{\IEEEauthorrefmark{1}
Both authors contributed equally to this work}
\IEEEauthorblockA{\IEEEauthorrefmark{2}
University of New South Wales, Australia}
\IEEEauthorblockA{\IEEEauthorrefmark{3}Data61, CSIRO, Australia Email: \{zhi.zhang,dongxi.liu,surya.nepal\}@data61.csiro.au}
\IEEEauthorblockA{\IEEEauthorrefmark{4}Baidu Security Email:chengyueqiang@baidu.com}
\IEEEauthorblockA{\IEEEauthorrefmark{5}Florida State University, America Email:zwang@cs.fsu.edu}
\IEEEauthorblockA{\IEEEauthorrefmark{6}University of Adelaide Email: yval@cs.adelaide.edu.au}}


\maketitle

\begin{abstract}
Rowhammer is a hardware vulnerability in DRAM memory, where repeated access to memory can induce bit flips in neighboring memory locations. Being a hardware vulnerability, rowhammer bypasses all of the system memory protection, allowing adversaries to compromise the integrity and confidentiality of data. Rowhammer attacks have shown to enable privilege escalation, sandbox escape, and cryptographic key disclosures.

Recently, several proposals suggest exploiting the spatial proximity between the accessed memory location and the location of the bit flip for a defense against rowhammer. These all aim to deny the attacker's permission to access memory locations near sensitive data.

In this paper, we question the core assumption underlying these defenses. We present \upperpt, a confused-deputy attack that causes accesses to memory locations that the attacker is not allowed to access. Specifically, \upperpt exploits the address translation process of modern processors, inducing the processor to generate frequent accesses to protected memory locations. We implement \upperpt, demonstrating that it is a viable attack, resulting in a system compromise (e.g., kernel privilege escalation). We further evaluate the effectiveness of proposed software-only defenses showing that \upperpt can overcome those.

\end{abstract}


%
\IEEEpeerreviewmaketitle

\section{Introduction}\label{sec:intro}
In 2014, Kim et al.~\cite{kim2014flipping} performed the first comprehensive
study of an infamous software-induced hardware fault, the so-called
\emph{rowhammer} vulnerability.  
Specifically, frequent accesses to the same addresses in two
DRAM (Dynamic Random Access Memory) rows (known as \emph{hammer rows}) can cause
bit flips in an adjacent row (the \emph{victim row}). 
If the victim row contains sensitive data, such as page tables, 
these bit flips can corrupt the data and compromise the security of the system.
Because the adversary does not access the victim row, rowhammer attacks
can be carried out even when the attacker has no access to the sensitive data.
Thus, rowhammer attacks can 
bypass MMU-based domain isolation both between processes and between user and kernel spaces,
even in the absence of software vulnerabilities.
Rowhammer attacks have been shown to allow 
privilege escalation~\cite{seaborn2015exploiting,
gruss2016rowhammer, bosman2016dedup, frigo2018grand,
tatar2018throwhammer,gruss2017another,van2016drammer, xiao2016one} and to steal
private data~\cite{razavi2016flip,bhattacharya2016curious, kwong2020rambleed}.

\begin{figure}
\centering
\includegraphics[width=\columnwidth]{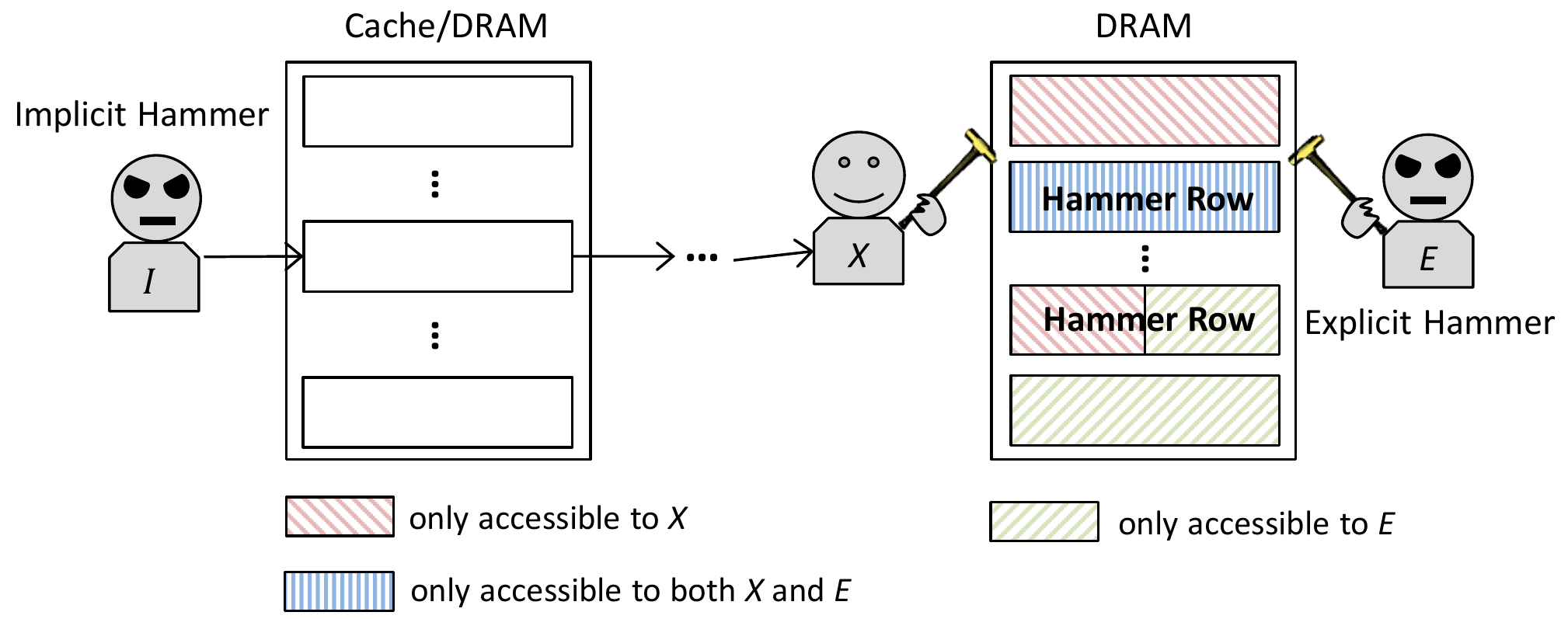}
\caption{
  Implicit\NCH and \upperperi. 
  \emph{All} existing rowhammer attacks are \upperperi, i.e.\ the attacker
  \emph{E} requires access to memory in  the exploitable hammer rows.
  In \uppertele, in contrast, the attacker \emph{I} 
  exploits a benign entity \emph{X} (e.g., the processor) to implicitly\NCH
  hammer the exploitable hammer rows.} 
\label{fig:indirect_hammer}
\end{figure}

One limitation of existing rowhammer attacks is that the adversary
requires access to an exploitable hammer row.
(A hammer row is \emph{exploitable} if the adversary can use it for the attack,
i.e., it is adjacent to a victim row that contains sensitive data.%
\footnote{
  In the RAMBleed attack~\cite{kwong2020rambleed}, the exploitable rows
  are the rows that contain the sensitive data.
})
That is, as \autoref{fig:indirect_hammer} shows,  some memory in the hammer row 
should be mapped to the 
address space of the attacker, who should have the permission
to read that memory.
As the access to the hammer row is
legitimate and conforms to the privilege boundary enforced by MMU, we refer to such
attacks as \emph{\upperperi}\NCH.

This limitation of \upperperi attacks underlies the design of \new{some} proposed 
software-only defenses against rowhammer~\cite{bock2019rip,brasser17can,wu2018CAT}.
These defenses enforce
DRAM-aware memory isolation at different granularities to deprive attackers
of access to exploitable hammer rows. As an example, CTA~\cite{wu2018CAT}
allocates memory for page tables in a separate region of the DRAM,
such that no row adjacent to page tables is accessible to unprivileged
users.
Unlike hardware-based defenses (e.g., ~\cite{DDR4, LPddr4,son2017making, lee2019twice}), 
such software-only defenses have the appeal of compatibility with 
existing hardware, which allows better deployability.

In this paper, we question the assumption underlying proposed software-only defenses.
Specifically, we ask:

\medskip
\begin{center}
  \emph{Are rowhammer attacks feasible without access permission to exploitable hammer rows?}
\end{center}
\medskip

\mypara{Our contributions}
In this paper, we provide a positive answer to this question.
We introduce a new class of rowhammer attacks, called
\emph{\uppertele}. As \autoref{fig:indirect_hammer} shows, an attacker \emph{I}
in \uppertele uses a confused deputy attack~\cite{Hardy88},
to bypass the access restrictions.
Instead of explicitly accessing the exploitable hammer rows,
the attacker tricks a benign entity to implicitly hammer the exploitable rows,
eliminating the key requirement of \upperperi.
Essentially, \uppertele \xspace \new{exploits} built-in features of modern hardware and/or software, such as address translation, a system call handler, etc.,
where the entity is either hardware, e.g.,\ the processor, or software, such as system call
handler. For instance, an unprivileged attacker can invoke a system call to cross user-kernel privilege boundary and access kernel
memory~\cite{lipp2018meltdown} implicitly. If such access frequently occurs within DRAM,
then the accessed kernel memory might be vulnerable to rowhammer.

\eat{Essentially, \uppertele abuses built-in features of modern hardware and/or software, such as speculative execution, a system call handler, etc.,
where the entity is either hardware, e.g.,\ the processor, or software, such as system call
handler. For instance, an unprivileged attacker can exploit out-of-order
execution to cross user-kernel privilege boundary and access kernel
memory~\cite{lipp2018meltdown}. If such access frequently occurs within DRAM,
then the accessed kernel memory might be vulnerable to rowhammer.
\yval{A concern with this paragraph is that it seems that we promise to demonstrate attacks through speculative execution and through system call handlers.}
}

Carrying out a \uppertele attack raises the following challenges as illustrated in
Figure~\ref{fig:indirect_hammer}.  First, the attacker \emph{I} should find 
a hardware or software feature that implicitly accesses the hammer rows
(solid arrows).  Second, the attacker \emph{I} should
effectively and efficiently trigger the selected feature to hammer the hammer
rows.  Third, the hammer rows (blue) should be exploitable for a
meaningful attack~\cite{kwong2020rambleed,brasser17can}. 
  
\mypara{\upperpt}
To demonstrate the viability of \uppertele, we instantiate a concrete example
that is called \emph{\upperpt}.  Specifically, like
van Schaik et al.~\cite{van2018malicious}, we use the page translation mechanism
of the processor as a confused deputy.
In modern mainstream operating systems, a memory access triggers
address translation. Specifically, when a program
accesses memory, the processor needs to translate the virtual address that
the program uses to a physical address.
On the Intel x86 architecture, the processor first searches the 
Translation-Lookaside Buffer (TLB) for the corresponding physical address.
In a TLB miss, when the search fails,
the processor proceeds to search the  paging-structure caches that store
partial address mappings of different page-table
levels~\cite{barr2010translation}. 
Finally, 
if no partial translation is found, the processor translates the address
by ``walking'' the page tables.
In the page-table walk, page-table entries (PTEs) are retrieved from
the cache, if they are cached, or from the DRAM memory, if not.

In \upperpt we exploit this page-table walk to perform an \uppertele.
Theoretically, all of the levels of PTEs
can be used for implicitly hammering memory\NCH. 
However, for that to happen, the attacker needs to ensure
that the corresponding entries are evicted from the TLB, the 
paging-structure cache, and the data caches.
At the same time, both evicting entries from caching structures
and page-table walks take time,
hence a naive implementation of the attack may not be fast
enough to induce bit flips.
Thus the major challenge of \upperpt is how to
exploit the page-table walk mechanism to produce frequent enough
memory accesses to exploitable hammer rows.

To address this challenge, exploit an interaction between the paging-structure
cache and page-table walks.
Specifically, if a partial translation for a page-table walk exists
in the cache, the processor uses the partial translation to skip 
parts of the page-table walk.
Thus, if we ensure that the page-table walk only misses on the Level-1 PTE (L1PTE), we can perform an implicit memory access to a single L1PTE only.

For \upperpt, we prepare a pool of eviction sets for the TLB and for the cache.
Each of these eviction sets allows us to evict one set of the TLB 
or the cache. 
We then repeatedly select a pair of memory addresses.
For each of these pairs, we repeatedly hammer the memory row
that contains the L1PTEs of the addresses.
To that aim, 
we use these eviction sets to evict the TLB sets 
that store the entries of the selected pair of addresses,
as well as the last-level cache sets that store the L1PTEs for
these addresses.
We then access the memory addresses.
Because we evicted the TLB entry, the processor needs to perform a
page-table walk.
Most of the address translation is cached in the paging-structure cache.
However, the entry of the L1PTE is not cached and retrieving it
requires a DRAM access.
If these L1PTEs happen to be in exploitable hammer rows, we can
expect hammering to induce a bit flip in the victim row.

We evaluate \upperpt in two system settings, when using regular (4\,KiB) pages,
and with huge (2\,MiB) \emph{superpages}. 
(As we show in \autoref{sec:eva}, the latter facilitates the faster generation
of eviction sets.)
The experimental results indicate that \upperpt is able to
cross the user-kernel boundary, allowing an unprivileged user 
to induce exploitable bit flips in L1PTEs and
to gain kernel privilege in either setting. 
We further show that \upperpt can overcome all of the 
aforementioned practical defenses (\autoref{sec:eva}).
To the best of our knowledge, we are the
first to demonstrate an attack capable of compromising the CTA defense~\cite{wu2018CAT}. 
We discuss other possible instances of \uppertele in Section~\ref{sec:dis}.

\mypara{Summary of Contributions}
The main contributions of this paper are as follows:
\begin{itemize}[noitemsep, topsep=2pt, partopsep=0pt,leftmargin=0.4cm]
    \item We demonstrate \upperpt, the first \uppertele attack, which 
      exploits page-table walks to void the
      core assumption that underlies all of the published 
      software-only defenses for rowhammer. (\autoref{sec:overview}.)

    \item We identify and exploit an efficient 
      page-table walk path that only
      induces loads of L1PTEs from memory, builds eviction sets to flush
      relevant hardware caches fast enough to cross the user-kernel boundary in
      hammering L1PTEs and gain kernel privilege. (\autoref{sec:overview}.)

    \item We evaluate \upperpt on three different machines, 
      in two system settings, with and without superpages,
      demonstrating privilege escalation with \uppertele. (\autoref{sec:eva}.)
    
    \item We evaluate multiple proposed software-only defenses~\cite{brasser17can,bock2019rip,wu2018CAT}, and show that \upperpt bypasses all of them. (\autoref{sec:defenses}.)
\end{itemize}

\section{Background and Related Work}\label{sec:bkgd}

In this section, we introduce the rowhammer bug and its attacks.

\eat{
\subsection{CPU Cache}\label{sec:cache}
In commodity Intel x86 micro-architecture platforms, there are three levels of CPU caches. Among all levels of caches, the first level of cache (i.e., L1 cache) is closest to CPU.
L1 cache is divided into two, the L1D caching data and the L1I caching instructions. The second level of cache, L2, is unified caching both data and instructions. Similar to the L2, the last-level cache (LLC) or L3, is also unified.  
Generally speaking, a cache of a specific level is set-associative and consists of \emph{S} sets. Each set contains \emph{L} lines and data or code can be cached in any line of the set; this is referred as an \emph{L}-way set-associative cache. Each cache line stores \emph{B} bytes. Thus, the overall cache size of that level is $\emph{S} \times \emph{L} \times \emph{B}$.  

To find data in the cache, the Intel micro-architectures use its virtual or physical address to decide its corresponding cache set of a specific cache level. For instance, L1 cache set is indexed using bits 6 to 11 of a virtual address. For L3, its indexing scheme is more complicated. In contrast to L1 and L2 that are private to a physical core, L3 is shared among all cores. So L3 cache is firstly partitioned into slices, and one slice serves one core with a faster access. Each slice is further divided into cache sets as described above. As such, some physical-address bits are XORed to decide a slice, and some bits (bits 6 to 16) are XORed to index a cache set~\cite{liu2015last}.

\subsection{Translation-Lookaside Buffer}
The Translation Lookaside Buffer (TLB) facilitates address  translation and has two levels. 
The first-level (i.e., L1), consists
of two parts: one that caches translations for code pages, called L1 instruction TLB (L1 iTLB), and the other that caches translations for data pages, called L1 data TLB (L1 dTLB). The second level TLB (L2 sTLB) is larger and shared for translations of both code and data.
Similar to the CPU cache above, the TLB at each level is also partitioned into sets of ways. One way is a TLB entry that stores one address mapping between a virtual address and a physical address~\cite{gras2018translation}. 

Note that a virtual address (VA) determines a TLB set of each level. Although there is no public information about the mapping between the VA and the TLB set, it has been reverse-engineered on quite a few Intel commodity platforms~\cite{gras2018translation}. 

\eat{
\subsection{Address Translation} \label{sec:page-table-walk}
Memory Management Unit (MMU) enforces memory virtualization primarily by the means of paging mechanism. Paging on the x86-64 platform usually uses four levels of page tables to translate a virtual address to a physical address. As such, virtual-address bits are divided into 4 parts. 
The bits 39$\sim$47 are used to index the Page Map Level table (the PML4 base address is in \texttt{$CR_3$}) and consequently these bits decide the page offset of a PML4 entry. The bits 30$\sim$38 are used to index a selected page directory pointer table (the base address of PDPT comes from the PML4 entry and the entry determines a physical page frame number and attributes (e.g., access rights) for access to the physical page).
The bits 21$\sim$29 are used to index a selected page directory (PD) table (the base address of PD comes from the PDPT entry). 
The bits 12$\sim$20 are used to index a selected page table (the base address of PT comes form the PD entry). Now the indexed PT entry points to the physical address of a corresponding page and the rest bits 0$\sim$11 are the offset into that page.

In order to facilitate the process, TLB is introduced to cache the address translations while cache is involved to store the accessed data as well as the page table entries of all levels.
}

\eat{
For the sake of performance, recent address translations
are aggressively cached in TLB, indicating that fresh translations might evict obsolete translations once they are mapped to the same TLB set. As the mapping between a virtual address and multi-level TLB sets has not been documented by Intel, it has been reverse engineered by Gras et al.~\cite{gras2018translation}, we can utilize the mapping to create a minimal eviction set to evict target address translations from TLB effectively and efficiently. 
}

\subsection{Dynamic Random-access Memory}
The Main memory of most modern computers uses Dynamic Random-access Memory (DRAM). Memory modules are usually produced in the form of dual inline memory module, or DIMM, where both sides of the memory module have separate electrical contacts for memory chips. Each memory module is directly connected to the CPU's memory controller through one of the two channels. Logically, each memory module consists of two ranks, corresponding to its two sides, and each rank consists of multiple banks. A bank is structured as arrays of memory cells with rows and columns. 

Every cell of a bank stores one bit of data whose value depends on whether the cell is electrically charged or not. A row is a basic unit for memory access. Each access to a bank ``opens'' a row by transferring the data from all of the cells of the row to the bank's \emph{row buffer} that acts as a cache for the most recently accessed row. This operation discharges all the cells of the row. To prevent data loss, the row buffer is then copied back into the cells, thus recharging the cells. Consecutive access to the same row is fulfilled from the row buffer,  while accessing another row flushes the row buffer. 
}

\subsection{Rowhammer Bug}\label{sec:hammeroverview}

Kim et al.~\cite{kim2014flipping} discovered that current DRAMs are vulnerable to disturbance errors induced by charge leakage. In particular,  their experiments have shown that frequently opening the same row  (i.e., hammering the row) can cause sufficient disturbance to a neighboring row and flip its bits without even accessing the neighboring row. Because the row buffer acts as a cache, another row in the same bank is accessed to flush the row buffer after each hammering so that the next hammering will re-open the hammered row, leading to bit flips of its neighboring row. 

\mypara{Hammering techniques}
Generally, there are three techniques for hammering a vulnerable DRAM.

\emph{Double-sided hammering:} is the most efficient technique to induce bit flips. Two adjacent rows of a victim row are hammered simultaneously and the adjacent rows are called \emph{hammer rows} or aggressor rows~\cite{kim2014flipping}. 


\emph{Single-sided hammering:}
Seaborn et al.~\cite{seaborn2015exploiting} proposed a single-sided hammering by randomly picking multiple addresses and hammering them with the hope that such addresses are in different rows within the same bank. 

\emph{One-location hammering:}
one-location hammering ~\cite{gruss2017another} randomly selects a single address for hammering. It exploits the fact that advanced DRAM controllers apply a sophisticated policy to optimize performance, preemptively closing accessed rows earlier than necessary. 

\mypara{Key requirements}
The following requirements are needed by \upperperi-based attacks to gain either privilege escalation or private information. 

\emph{First}, 
the CPU cache must be either flushed or bypassed. 
It can be invalidated by instructions such as \texttt{clflush} on x86. In addition, conflicts in the cache can evict data from the cache since the cache is much smaller than the main memory. Therefore, to evict hammer rows from the cache, we can use a crafted access pattern~\cite{rowhammerjs} to cause cache conflicts for  hammer rows. Also, we can bypass the cache by accessing uncached memory.

\emph{Second}, the row buffer must be cleared between consecutive hammering DRAM rows. Both double-sided and single-sided hammering explicitly perform alternate access to two or more rows within the same bank to clear the row buffer. 
One-location hammering relies on the memory controller to clear the row buffer.

\emph{Third}, existing rowhammer attacks require that at least part of a hammer row be accessible to an attacker in order to gain the privilege escalation or steal the private data, such that a victim row can be compromised by hammering the hammer row. 

\emph{Fourth}, either the hammer row or the victim row must contain sensitive data objects (e.g., page tables) we target. 
If the victim row hosts the data objects, an attacker can either gain the privilege escalation or steal the private data~\cite{seaborn2015exploiting,bhattacharya2016curious}. If the hammer row hosts the data objects, an attacker can steal the private data~\cite{kwong2020rambleed}.

\eat{
Usually, the attacker does not have direct control of the (physical) memory allocation. To address that, the attacker will spray the memory with numerous copies of 
a probabilistic approach is usually adopted on the x86 architectures. Specifically, the attacker allocates a large number of potential hammer rows and induces the kernel to create many copies of the target objects. This strategy is very similar to the heap spray attack in that by spraying the memory with potential hammer and victim rows, the probability of the correct positioning is high. Page tables are often targeted as the victim row because they control the system memory mapping and it is relatively easy to create many page-table pages (by allocating and using a large block of memory). An attacker-controlled page table essentially allows him to read/write/execute all the memory in the system. 
}

\subsection{Rowhammer Attacks}\label{sec:related}
In order to trigger rowhammer bug, frequent and direct memory access is a prerequisite. Thus, we classify rowhammer attacks into three categories based on how they flush or bypass cache. 

\mypara{Instruction-based}
Either \texttt{clflush} or \texttt{clflushopt}  instruction is commonly used for explicit cache flush~\cite{kim2014flipping, seaborn2015exploiting,gruss2017another,razavi2016flip,cojocar2019exploiting,cojocar2020we,frigo_trrespass_2020} ever since Kim et al.~\cite{kim2014flipping} revealed the rowhammer bug. 
Also, Qiao et al.~\cite{qiao2016new} reported that non-temporal store instructions such as \texttt{movnti} and \texttt{movntdqa} can be used to bypass cache and access memory directly. 


\mypara{Eviction-based}
Alternatively, an attacker can evict a target address by accessing congruent memory addresses which are mapped to the same cache set and same cache slice as the target address~\cite{aweke2016anvil, rowhammerjs, bosman2016dedup,liu2015last,maurice2017hello,van2018malicious}. A large enough set of congruent memory addresses is called an eviction set. Our \upperpt also chooses the eviction-based approach to evict Level-1 PTEs from cache. 

\mypara{Uncached Memory-based}
As direct memory access (DMA) memory is uncached, past rowhammer attacks such as Throwhammer~\cite{tatar2018throwhammer} and Nethammer~\cite{lipp2018nethammer} on x86 microarchitecture and Drammer~\cite{van2016drammer} on ARM platform have \new{leveraged} DMA memory for hammering. Note that such attacks hammer target rows that are within an attacker's access permission. 



\section{Overview}\label{sec:overview}
In this section, we first present the threat model and assumptions, and then discuss \upperpt in detail. 

\subsection{Threat Model and Assumptions}
Similar to previous rowhammer attacks~\cite{xiao2016one, razavi2016flip,
qiao2016new, bosman2016dedup, gruss2016rowhammer, seaborn2015exploiting},
we assume an unprivileged attacker that tries to cause a bit flip in sensitive data
that the attacker is not allowed to access, let alone modify.
We further assume that the attacker does not know the location of the sensitive
data, i.e., its physical address, and does not have access to interfaces,
such as \texttt{pagemap}, that convert between virtual and physical
addresses.
Additionally, we assume that the software and the operating system 
are working correctly and have no software vulnerabilities.
Last, like prior attacks, we assume that the memory is vulnerable to
the rowhammer attack.
Pessl et al.~\cite{pessl2016drama} report that many DRAM modules,
including both DDR3 and DDR4 modules, sold by 
mainstream DRAM manufacturers are vulnerable.

Unlike past works, we assume that the system is protected by software-only
defenses, such as RIP-RH~\cite{bock2019rip}, CATT~\cite{brasser17can},
or CTA~\cite{wu2018CAT}. These defenses segregate the sensitive data in
memory to prevent attacker's access to exploitable hammer rows.

\subsection{\upperpt}\label{sec:pthammer}
\upperpt is page-table-based \uppertele attack that leverages the address translation
feature of the processor to hammer Level-1 page tables (L1PTs) implicitly 
and to flip exploitable
bits in other L1PTs, thereby achieving privilege escalation. 
Specifically, we observe that the address translation feature enables 
implicit access to the page tables, which are served from the DRAM memory. 
Based on this observation, we build an attack primitive that
hammers Level-1 page-table entries (L1PTEs). Using this hammer primitive, we
induce bit flips in sensitive data in the kernel.
In particular, in our implementation we induce bit flips in L1PTEs,
to compromise them and gain kernel privilege.

At a high level, \upperpt follows in the footsteps of the 
``Malicious Management Unit'' attack of van Schaik et al.~\cite{van2018malicious}.
Like their attack, \upperpt is a confused-deputy attack that exploits
the memory management unit to bypass memory-segregation defenses.
However, unlike van Schaik et al., we do not care about cache noise,
but have tight timing constraints required for achieving bit flips.
Thus, our focus is on performing the attack \emph{efficiently}
and \emph{effectively}.

\mypara{Address Translation in Intel x86}
Modern operating systems isolate user processes by running each user
process in a \emph{virtual address space}.  
The operating system and the MMU collaborate on translating the
virtual addresses that processes use to \emph{physical addresses},
which determine the location of the data in the memory.
The main data structure used for this translation is the \emph{page tables},
This is a four-level tree where each level is indexed by 9 address bits,
covering a virtual address space of 48 bits.
To translate an address, the MMU performs a \emph{page-table walk}
querying the page tables from the root of the tree down to the lower
Level 1, which contains the translation of the address.

\begin{figure}
\centering
\includegraphics[scale=0.45]{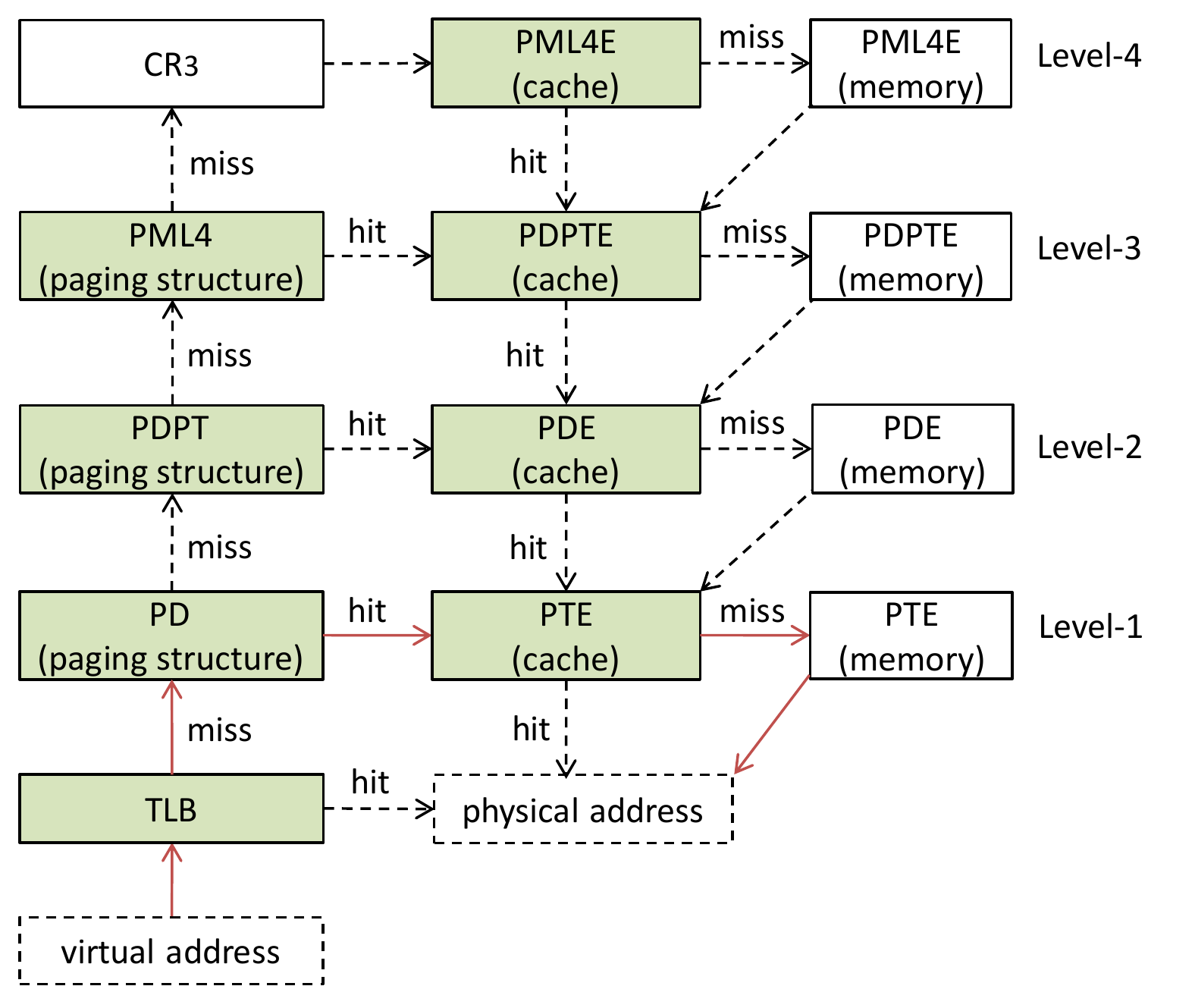}
\caption{Address translation in Intel x86. Red solid
  arrows mark the path that \upperpt uses to implicitly
  access a Level-1 page-table entry (PTE) from memory. 
  To that aim,
  \upperpt flushes the TLB entry and the cached PTE for the target address,
  while retaining all of the higher-level partial translations in the
  respective paging-structure caches.
  }
\label{fig:pt_walk}
\end{figure}

To reduce the cost of page-table walks, the MMU also caches prior
translation results, which are then used to bypass parts or all
of the page-table walk.
As \autoref{fig:pt_walk} shows,  the MMU maintains a separate caching
structure for each of the levels of the page tables.
The Translation-Lookaside Buffer (TLB) caches complete translations.
The other \emph{paging-structure caches} cache partial 
translations~\cite{barr2010translation}.
Thus, when translating a virtual address,
the MMU first checks for the corresponding physical address in the TLB.
If the address is not in the TLB, it proceeds to search the PD, which
caches location of level 1 page tables.
The search proceeds up the hierarchy, until a page is found
or the MMU exhausts the paging-structure caches.
The MMU then performs a page-table walk from either the
cached result or from the root of the page-tables tree.

The page tables themselves are stored in the DRAM. When accessing
them during the page-table walk, the MMU first searches
for the required page-table entry in the data caches,
and accesses the DRAM only if no cached copy is found in the data
cache hierarchy.


\mypara{An Implicit Memory Access Primitive}
For \upperpt we want to \new{exploit} the address translation, and specifically the page-table
walk, to achieve \uppertele. 
This raises two requirements.  First, we need to perform actual DRAM accesses,
and second we need to performs these accesses fast enough.
However, these requirements conflict, because DRAM accesses take time and
to perform these we need to evict information from various caching structures
which require further time.
Hence the challenge is to ensure that DRAM accesses happen without spending
too much time.

To address this challenge, we identify the shortest path through address
translation that results in a memory access.
This path is highlighted with solid red arrows in \autoref{fig:pt_walk}.
As we can see in the diagram, to traverse this path we need address
translation to miss on the TLB, hit on the PD, and miss on the PTE access
in the cache.
Thus, given a target address whose L1PTE entry resides in an exploitable
hammer row, to induce an implicit access to the exploitable hammer row,
we need to evict the entry of the target address from the TLB and evict
L1PTE entry for the target address from the data cache.

\mypara{Finding Exploitable Target Addresses}
To complete the attack, we need to find a victim row that has a flippable bit
and contains sensitive data.
We further need to implicitly hammer the two adjacent rows.
That is, we need to find two target memory addresses whose
L1PTEs are in the rows adjacent to the victim row.
We achieve  this using memory spraying~\cite{cheng2018still, seaborn2015exploiting}.
We allure the kernel to create a large number of L1PT pages, 
filling a significant part of the memory with such pages.
In such an arrangement, a  random bit flip in memory has a 
non-negligible probability of changing 
the address of one of the L1PTEs.
Because a significant part of the memory contains L1PT pages,
there is a non-negligible probability that modified L1PTE
points to an L1PT page.
This, effectively, gives the attacker write access to an L1PT page.
Modifying this page, the attacker can get access to any desired physical 
page.

\mypara{Evicting the TLB entry and the L1PTE}
While the Intel architecture supports instructions for evicting entries
from the TLB and the cache, the former is restricted to privileged code
and the latter only works on data the user can access.
As such, we have to resort to causing contention on these components
to evict the entries.
We now discuss how we create eviction sets that allow effective
and efficient eviction of these entries.

\subsection{Evicting the TLB Entry}\label{sec:tlbflush}
As Gras et al.~\cite{gras2018translation} have revealed there exists an explicit mapping between a virtual page number and a multi-level TLB set, we simply create an initial eviction set that contains multiple (physical) pages to flush a cached virtual address from TLB. One subset of the pages is congruent and mapped to a same L1 dTLB set while the other is congruent and mapped to a same L2 sTLB set if TLB applies a non-inclusive policy. 

Take one of our test machines, Lenovo T420, as an example, both L1 dTLB and L2 sTLB have a $4$-way set-associative for every TLB set and thus $8$ (physical) pages are enough as an minimum eviction set to evict a target virtual address from TLB. However, when we create such an eviction set and  profile the access latency of a target virtual address, its latency remains unstable. 
To collect the number of TLB misses induced by the target address, we develop a kernel module that applies Intel Performance Monitoring Counters (PMCs) to count TLB-miss events (i.e., \texttt{dtlb\_load\_misses.miss\_causes\_a\_walk}). The experimental results show that TLB misses in both levels do not always occur when profiling the target address, meaning that the target address has not been effectively evicted by the eviction set, and thereby rendering our TLB flushes ineffective. A possible reason is that the eviction policy on TLB is not true Least Recently Used (LRU). 

\mypara{Decide a Minimal Eviction-Set Size for TLB}
To this end, we propose a working \autoref{alg:tlb_flush} that decides a
minimal size without knowing its eviction policy. Note that the minimal size is
used to prepare a minimal TLB eviction set in \upperpt.
Specifically, Lines 2--13 define a function $\pts$ that reports the TLB
miss probability
induced by accessing $\targetaddr$. Specifically, the function
accesses all the elements in the eviction set $\set$ (Lines 6--8)
aiming to evict the cached address
mapping for $\targetaddr$ from the TLB.
It counts the number of misses (Lines 9--11) and returns the ratio
of misses to tries (Line~13).
The main code starts 
from a large buffer $\buf$. In Lines 15--19, We select all those
pages that are indexed to the same TLB set as the $\targetaddr$ by leveraging
the reverse-engineered mapping function of Gras et al.~\cite{gras2018translation}.
Note
that the size of $\buf$ is determined based on the number of entries for 4\,KiB pages
in the TLB. 
If $\targetaddr$ is allocated from a huge page, 
the number of TLB entries that for the page size should be considered.  The
selected pages are then populated and inserted to $\initset$ (Lines 15--19).
Populate the selected pages is essential in order to trigger the
address-translation feature so tha the  TLB will cache address mappings
accordingly.  In Line~21, we find a threshold for effective TLB flushes.  We
then start to trim the set while retaining its effectiveness in
Lines 22--28.

\begin{algorithm}[t]
\small
	\caption{Find the  minimal eviction-set size for TLB }\label{alg:tlb_flush}
	\textbf{Initially:} $\targetaddr$ is a page-aligned virtual address
	that needs its cached TLB entry flushed. A buffer ($\buf$) is
	pre-allocated, size of which is decided by TLB entries. $\initset$ is an empty set. Two
	unique unsigned integers are assigned to $\datamarker$ and $\loopnum$,
	respectively.\\
	\SetKwProg{Fn}{Function}{}{}
	    \Fn {$\pts (\set)$} {
	        $\misses \leftarrow 0$ \\
	        \RepTimes{$\loopnum$} {
	            access $\targetaddr$ \\
	            \ForEach {$\page  \in \set$} {
	                access $\page[0]$ \\
	            }
		    \If{\emph{TLB miss when accessing} $\targetaddr$}{
		      $\misses \leftarrow misses + 1$
		    }
	        }
	       \KwRet $misses/\loopnum$ \\
        }
	   $\targetaddr \leftarrow \datamarker$ \\
	   \ForEach {$\page \in \buf$} {
	     \If {$\page$ \emph{and} $\targetaddr$ \emph{are in the same set}} {
            $\page[0]   \leftarrow \datamarker$ \\
            add $\page$ into $\initset$. \\
        }
      }  
	   $\threshold   \leftarrow \pts(\initset)$ \\
	    
	   \For{$\page \in \initset$} {
	       take one $\page$ out of $\initset$. \\
	       $\temptlbmiss  \leftarrow \pts(\initset)$ \\
	    \If {$\temptlbmiss < \threshold$} {
	        put $\page$ back into $\initset$ and break. \\
	    }
	   }
	   \KwRet the size of $\initset$ \\
\end{algorithm}

\subsection{Evicting the L1PTE from the Cache}\label{sec:cacheflush}
Now we are going to flush a cached Level-1 PTE (L1PTE) that corresponds to a
target virtual address.  Considering that last-level cache (LLC) is
inclusive~\cite{intelOp}, we target flushing the L1PTE from LLC such that the
L1PTE will also be flushed out from both L1 and L2 caches (we thus use cache
and LLC interchangeably in the following section).  In contrast to the TLB that
is addressed by a virtual page-frame number, the LLC is indexed by
physical-address bits, the mapping between them has also been reverse
engineered~\cite{hund2013practical,Maurice2015,irazoqui2015systematic}. Based
on the mapping, we decide the size of a minimal LLC eviction set in an offline
phase where physical addresses are available. When launching \upperpt, we build
a one-off  pool of minimal eviction sets for every LLC set and select one from
the pool for a target L1PTE.  In the following, we talk about the above three
steps in detail. 

\mypara{Decide a Minimal Eviction-Set Size for LLC}
We extend the aforementioned kernel module to count the event of LLC misses
(i.e., \texttt{longest\_lat\_cache.miss}) and have a similar algorithm to
\autoref{alg:tlb_flush} to decide the minimal size for an LLC eviction
set, namely, prepare a large enough eviction set congruent as a target virtual
address and gain a threshold of LLC-miss number induced by accessing the target
address, remove memory lines randomly from the set one by one and verify
whether currently induced LLC-miss number is less than the threshold. If yes, a
minimal size is determined. Also, this algorithm is performed in an offline
phase long before \upperpt is launched. 

Although the size of eviction-set is determined ahead of time, \upperpt in our
threat model cannot know the mapping between a virtual and a physical address,
making it challenging to prepare an eviction set for any target virtual address
during its execution. Also, \upperpt cannot obtain the L1PTE's physical
address, and thus it is difficult to learn the L1PTE's exact location (e.g.,
cache set and cache slice) in LLC.  To address the above two problems, \upperpt
at the  beginning prepares a complete pool of eviction sets, which is used to
flush any target data object including the L1PTE. It then selects an eviction
set from the pool to evict a target L1PTE without knowing the L1PTE's cache
location. Note that preparing the eviction pool is a one-off cost and \upperpt
only need to repeatedly select eviction sets from the pool when hammering
L1PTEs.

\mypara{Prepare a Complete Pool of LLC Eviction Sets}
The pool has a large enough number of eviction sets and each is used to flush a
memory line from a specific cache set within a cache slice in LLC. The size of
each eviction set is the pre-determined minimum size. We implement the
preparation based on previous works~\cite{liu2015last,genkin2018drive}. Both
works  rely on the observation that a program can determine whether a target
line is cached or not by profiling its access latency. If a candidate set of
memory lines is its eviction set, then the target line's access latency is
above a time threshold after iterating all the memory lines within the
candidate set. 

Specifically, if a target system enables superpags, a virtual address
and its corresponding physical address have the same least significant 21 bits,
indicating that if we know a virtual address from a pre-allocated super page,
then its physical address bits 0--20 are leaked and thus we know the cache
set index that the virtual address maps to~\cite{liu2015last}. The only
unsolved is the cache slice index. Based on a past
algorithm~\cite{liu2015last}, we allocate a large enough memory buffer (e.g.,
twice the size of LLC), select memory lines from the buffer that have the same
cache-set index and group them into different eviction sets, each for one cache
slice. 

If superpages are disabled, then only the least significant 12 bits (i.e.,
4\,KiB-page offset) is shared between virtual and physical addresses and
consequently we know bits 6--11 of the  cache-set index. As such,
we utilize another previous work~\cite{genkin2018drive} to group potentially
congruent memory lines into a complete pool of individual eviction sets.
Compared to the above grouping operation, this grouping process is relatively
slower, since there are many more memory lines sharing the same partial
cache-set bits rather than complete bits.

\begin{algorithm}[t]
	\caption{Select a minimal LLC eviction set }\label{alg:cache_flush}
	\small
	\SetAlgoLined
	\textbf{Initially:} a virtual page-aligned address ($\targetaddr$) is
	allocated and needs its L1PTE cache-line flushed. A complete pool of
	individual eviction sets ($\evictionsets$). $\pteoffset$ is decided
	by $\targetaddr$. A unique unsigned number is assigned to $\loopnum$
	and a set ($\latencyset$ is initialized to empty).
	$\maxlatency$ is initialized to $0$ and indicates the maximum latency
	induced by accessing $\targetaddr$. $\maxset$ represents the eviction
	set used for the L1PTE cache flush. \\
		\SetKwProg{Fn}{Function}{}{}
	    \Fn {$\pes (\set, \target)$} {
	        \RepTimes{$\loopnum$} {
	            \ForEach {$\memoryline \in \set$} {
	                access $\memoryline$.
	            }
	            flush a target TLB entry. \\
	            $\latency$ is decided by accessing $\target$. \\
	            add $\latency$ to $\latencyset$ \\
	        }
	       \KwRet the median of $\latencyset$ \\
        }
	\ForEach {$\set \in \evictionsets$} {
	      obtain $\pageoffset$ from first memory line in $\set$.\\
	      \If {$\pageoffset == \pteoffset$} {
	         $\latency \leftarrow \pes(\set, \targetaddr)$. \\
	         \If {$\maxlatency < \latency$} {
	            $\maxlatency \leftarrow \latency$. \\
	            $\maxset \leftarrow \set$.\\
	         }
	      }
	     
	}
	\KwRet $\maxset$ \\
\end{algorithm}

\begin{table*}
\centering
\begin{tabular}{@{}llllll@{}}
\toprule
\multirow{2}{*}{\textbf{Machine}} & \multirow{2}{*}{\textbf{Architecture}} & \multicolumn{3}{c}{\textbf{CPU}} & 
\multirow{2}{*}{\textbf{DRAM}} \\ \cmidrule{3-5}
 &  &  {Model} & {TLB Assoc.} & {LLC Assoc. \& Size} &  \\ \midrule
{Lenovo T420} & {SandyBridge} & {i5-2540M} & {4-way L1d, 4-way L2s} & {12-way, 3\,MiB} & {8\,GiB Samsung DDR3} \\

{Lenovo X230} & 
{IvyBridge} & {i5-3230M} & {4-way L1d, 4-way L2s} & {12-way, 3\,MiB} & {8\,GiB Samsung DDR3} \\
{Dell E6420} & 
{SandyBridge} & 
{i7-2640M} & {4-way L1d, 4-way L2s}  &
{16-way, 4\,MiB} & {8\,GiB Samsung DDR3} \\
\bottomrule
\end{tabular}
\caption{System Configurations.}
\label{tab:config}
\end{table*}

\mypara{Select a Target LLC Eviction Set}
Based on the pool preparation, we develop an \autoref{alg:cache_flush} to
select an eviction set from the pool and evict a L1PTE corresponding to a
target address.

In Line 12, we enumerate all the eviction sets in the pool and collect those
sets that have the same page offset as the L1PTE in Line 14.  This collection
policy is based on an interesting property of the cache. Oren et
al.~\cite{oren2015spy} report that if there are two different physical memory
pages that their first memory lines are mapped to the same cache set of LLC,
then the rest memory lines of the two pages also share (different) cache sets.
This means if we request many (physical) memory lines that have the same page
offset as the L1PTE and access each memory line, then we flush the L1PTE from
LLC.

Lines 15--19 select the target eviction set from the collected ones.  In Line~15,
we profile every collected eviction set through a predefined function in
Lines 2--11. Within this function, we perform access to each memory line of one
eviction set, which will implicitly flush the L1PTE from LLC if the eviction
set is congruent with the L1PTE, and then flush the target TLB entry related to
$\targetaddr$ to make sure the subsequent address translation will access the
L1PTE. At last, we measure the latency induced by accessing $\targetaddr$.
Based on this function, we find the targeted eviction set that causes the
maximum latency in Lines 17--18, as fetching the L1PTE from DRAM is
time-consuming when accessing $\targetaddr$ triggers the address translation.
Give that LLC is shared between page-table entries and user data, we must
carefully set $\targetaddr$ to page-aligned (normally 4\,KiB-aligned) but not
superpage-aligned (normally 2\,MiB-aligned), that is, its page offset is
$0$ and different from $\pteoffset$, which is the page offset of L1PTE. As
such, they are placed into different cache sets and the selected eviction set
is ensured to flush the target L1PTE rather than $\targetaddr$. 

\section{Evaluation}\label{sec:eva}
We now turn evaluate \upperpt on three different hardware, summarized in 
\autoref{tab:config}, all running Ubuntu 16.04.
We test \upperpt both in the default memory configuration and with huge memory
pages (superpages) enabled.

\begin{figure*}[ht]
	\centering
	\begin{minipage}[b]{\columnwidth}
		\centering
		\includegraphics[scale=0.5]{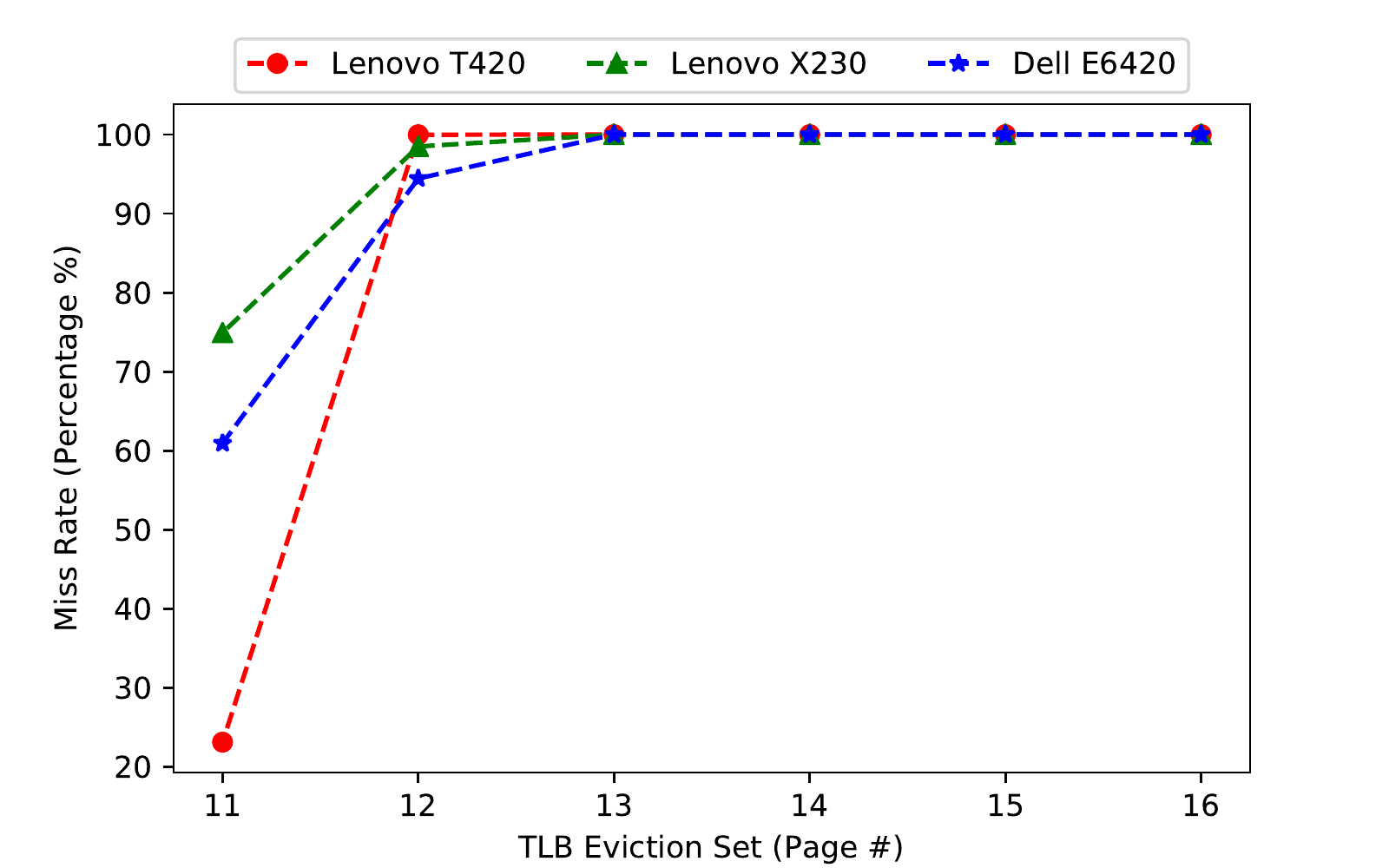}
		\caption{TLB miss rate for eviction set size. 
		  The TLB miss rate drops when using an eviction set of size below 12.}
		\label{fig:tlbflush}
	\end{minipage}
	\hfill
	\begin{minipage}[b]{\columnwidth}
		\centering
		\includegraphics[scale=0.5]{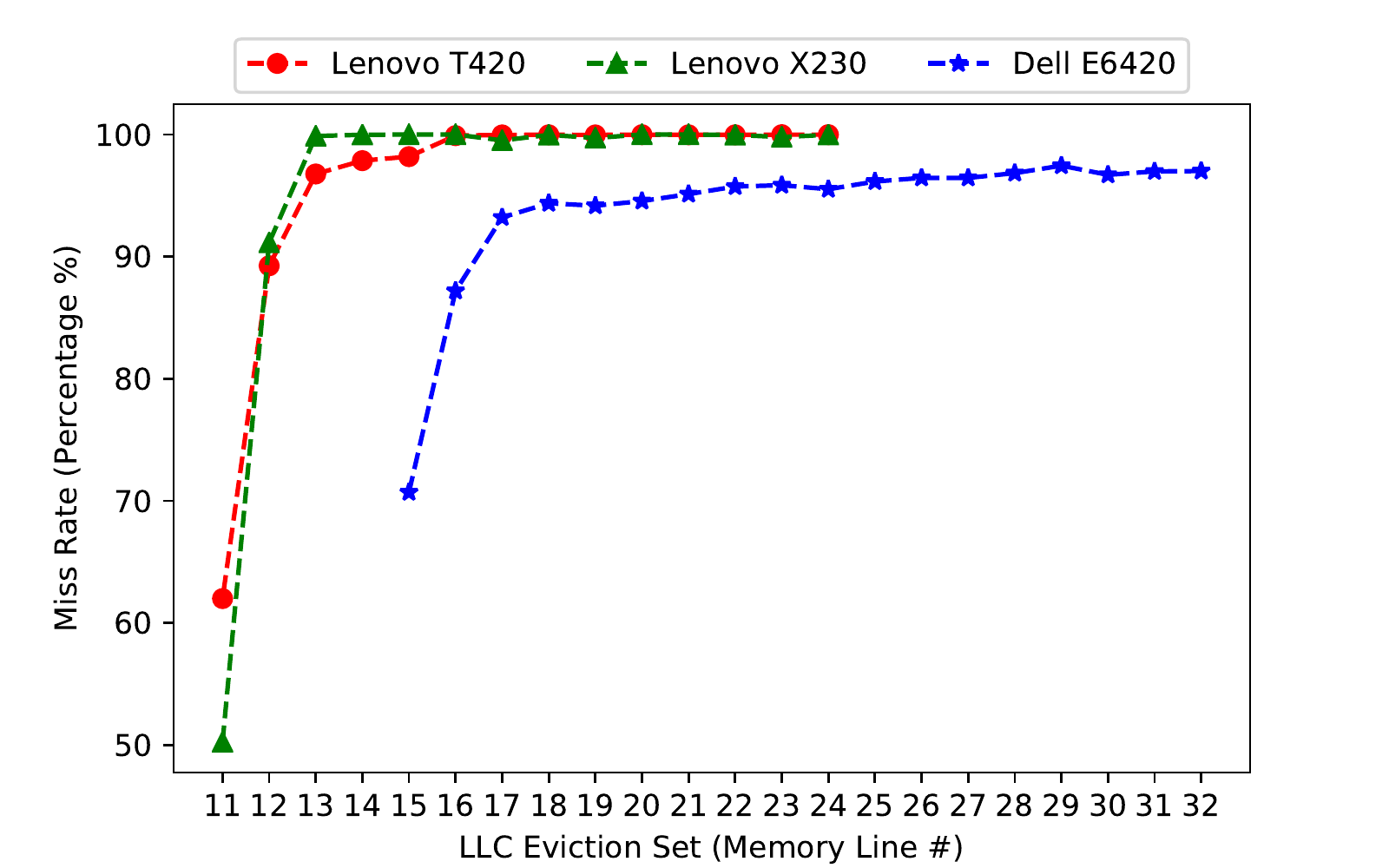}
		\caption{LLC miss rate for eviction set size.
		  When the size of the eviction set is larger than the LLC associativity,
		  the miss rate is consistently above 95\%.}
		\label{fig:cacheflush}
	\end{minipage}%
\end{figure*}

\begin{table*}
\centering
\begin{tabular}{llrrlllll}
\toprule
\multirow{2}{*}{\textbf{Machine}} & 
  \multirow{2}{*}{\textbf{Page Size}} &
  \multicolumn{2}{c}{\textbf{Preparation}} & 
  \multicolumn{2}{c}{\textbf{{Set Selection}}} & 
  \textbf{Hammer} &
  \textbf{Check} &
  \textbf{Time to}  \\
& 
  & 
  \multicolumn{1}{l}{TLB} & 
  \multicolumn{1}{l}{LLC} & 
  \multicolumn{1}{l}{TLB} & 
  \multicolumn{1}{l}{LLC} & 
  \textbf{Time} &
  \textbf{Time} &
  \textbf{Bit Flip}  \\ 
\midrule
\multirow{2}{*}{Lenovo T420} & superpage                  & 11\,ms &  0.3\,m & 1\,\us & 285\,ms & 285\,ms & 4.4\,s & 10\,m \\
			     & regular                    & 11\,ms & 18.0\,m & 1\,\us & 283\,ms &   287\,ms  &   4.4\,s     & 10\,m \\
\multirow{2}{*}{Lenovo X230} & \rule{0pt}{2.3ex}superpage &  7\,ms &  0.3\,m & 1\,\us & 282\,ms &   280\,ms      &  4.4\,s      & 15\,m \\
			     & regular                    &  7\,ms & 19.0\,m & 1\,\us & 288\,ms &    283\,ms     &  4.2\,s      & 15\,m \\
\multirow{2}{*}{Dell E6420}  & \rule{0pt}{2.3ex}superpage &  7\,ms &  0.3\,m & 1\,\us & 258\,ms &  389\,ms       &  4.1\,s  & 14\,m \\
			     & regular                    &  7\,ms & 38.0\,m & 1\,\us & 270\,ms &    392\,ms     &  4.0\,s      & 12\,m \\
 \bottomrule
\end{tabular}
\caption{Average time for \upperpt (five runs). First bit flip observed within 15 minutes of double-sided hammering.
Pool preparation is a one-off cost, accrued only once at the beginning of the attack.
Hammer and check times are the time it takes to perform a hammering attempt and to check
for bit flips, respectively.
}
\label{tab:timecosts}
\end{table*}
 
We first decide the minimal eviction-set size to effectively and efficiently
flush the TLB and the last-level cache (LLC) at an offline stage. Based on the minimal
size, we prepare a minimal TLB or LLC eviction sets from a complete pool of TLB
or LLC eviction sets.
We then evaluate the performance of the complete attack,
describe how it achieves privilege escalation,
and explore its effectiveness against proposed defenses.

\subsection{Eviction-Set Size}
\mypara{TLB} 
We use \autoref{alg:tlb_flush} (\autoref{sec:tlbflush}) to determine
the minimal eviction set size that consistently evicts an entry from 
the TLB. 
We first 
use the mapping of Gras et al.~\cite{gras2018translation} obtain an
initial eviction set twice bigger than the total associativity of the TLBs,
i.e., with 4-way L1dTLB and L2sTLB the initial eviction set has 16 elements.
We then measure the eviction success while reducing the eviction set size.
The results, presented in \autoref{fig:tlbflush},
show that in all of our test machines, eviction sets of 12 or more entries
achieve consistently high eviction rates,
while for smaller eviction sets the success drops significantly.

\mypara{LLC}
The associativity of the LLC varies between our test machines,
with the Lenovo machines having 12-way LLCs and the Dell machine using a
16-way LLC.
We use the algorithm of Liu et al.~\cite{liu2015last} to find conflicting
memory addresses and select initial eviction sets twice larger than 
a cache set, i.e., 24 entries for the Lenovos and 32 entries for the Dell.
\autoref{fig:cacheflush} shows the measured eviction rate while removing
elements from the eviction set.
We see that when the eviction set is bigger than the LLC set,
the eviction rate is consistently above 94\%.
However, the eviction rate starts dropping when the eviction set size
matches the cache associativity.
Further reducing the eviction set size results in a significant drop
in the eviction rate.
Thus, we choose an eviction set one larger than the cache associativity,
with 13 entries on the Lenovo machines and 17 on the Dell machine.
We note that Gruss et al.~\cite{rowhammerjs} explore the effects of
the order of access to the eviction set on the eviction rate.
We do not use their results, as we find that sequential access
produces sufficiently high eviction rates.

\subsection{Eviction Pool Preparation}
For TLB, we allocate a complete pool of 4\,KiB pages eight
times as many as required to cover both the L1dTLB and the L2sTLB entries 
for 4\,KiB-page.  We partition these pages based on the TLB sets they map to.
As \autoref{tab:timecosts} shows, this TLB pool preparation
completes within a few milliseconds on each of our test machines.

For the LLC, we allocate a buffer twice the LLC size and
use the algorithm of Liu et al.~\cite{liu2015last} to
partition it to  eviction sets.
Because the mapping of virtual to physical addresses preserves more
bits when using superpages, the pool preparation is significantly
faster when we use superpages.
The complexity of the algorithm we use is cubic with the size of the LLC.
Hence the algorithm is significantly slower on the Dell machine,
which features a larger cache.
As this is a one-off cost, we have not experimented with
potentially more efficient algorithms for 
finding cache sets, such as the Vila et al.~\cite{VilaKM19}.

\subsection{LLC Eviction-Set Selection}
TLB eviction-set selection relies on a complete reverse-engineered mapping
between virtual addresses and TLB sets~\cite{gras2018translation}, and thus it
introduces no false positives, meaning that \upperpt always selects a matching
eviction set for TLB.

However, selecting an LLC eviction set is based on profiling the access latency
to a target address, described in \autoref{alg:cache_flush}. As such, the
profiled latency is not completely precise due to noise, e.g., due to interrupts, and
may introduce false positives to the selection. 
To test the success of the algorithm, we develop a
kernel module that obtains the physical address of each L1PTE, which we use
to verify that the L1PTE is congruent with the eviction-set selected 
by \autoref{alg:cache_flush}
The experimental results show that the eviction-set selection for the LLC has no
more than 6\% false positives in each system setting on each test machine. 
The kernel module of SGX-Step~\cite{BulckPS17} can also be used to
find the physical address of the L1PTE.
\new{We note that this kernel module is \emph{not} required for the 
attack and is only used for evaluating the success of the eviction set selection.}

Note that selecting a TLB-based eviction-set takes about 1 microsecond while the
LLC eviction-set selection takes about 290 milliseconds. Both are quite
efficient, indicating that we can quickly start double-sided hammering, as
mentioned below. 

\begin{figure}
\centering
    \includegraphics[width=\columnwidth]{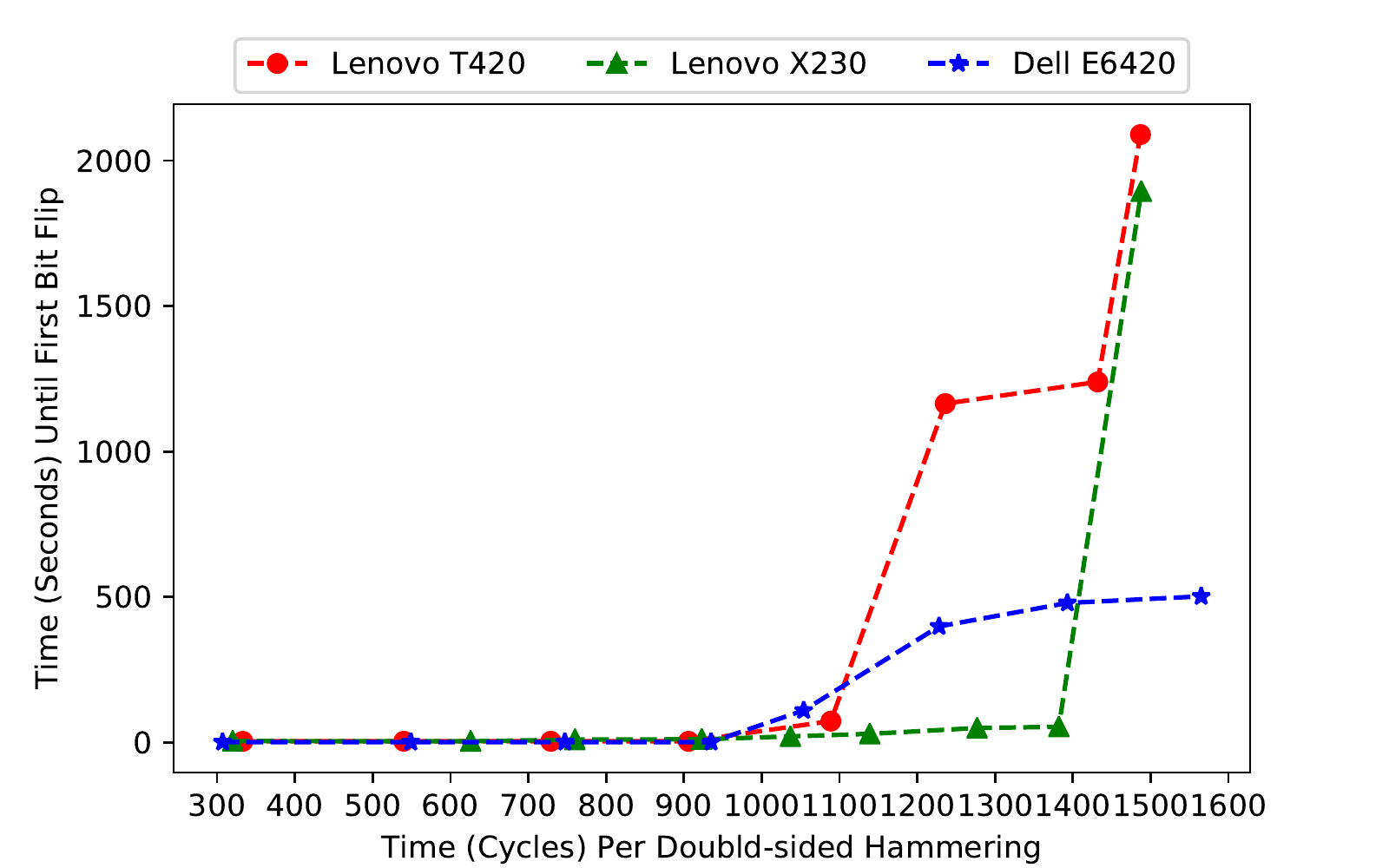}
\caption{As the time to perform an iteration of double-sided hammering
  increases, the time to find the first bit flip also grows. When 
  hammering iterations are longer than around \new{1\,600}
  cycles, no bit flip is observed within two hours.} 
\label{fig:distribution}
\end{figure}

\begin{figure*}
	\centering
	\begin{subfigure}[t]{\columnwidth}
		\centering
		\includegraphics[scale=0.5]{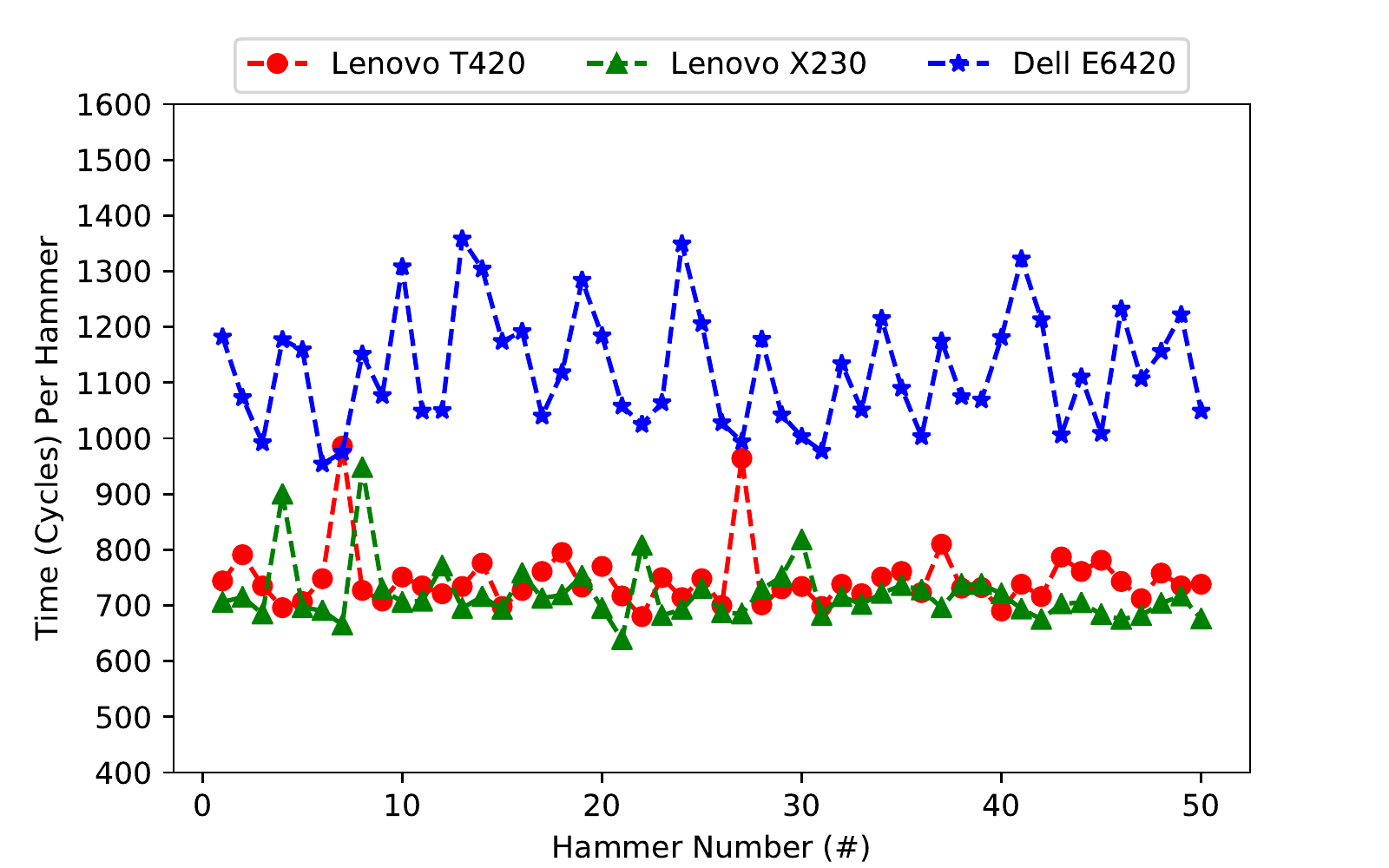}
		\caption{Double-sided hammering in the default memory setting.}
		\label{fig:4kbcyclerange}
	\end{subfigure}
	\hfill
	\begin{subfigure}[t]{\columnwidth}
		\centering
		\includegraphics[scale=0.5]{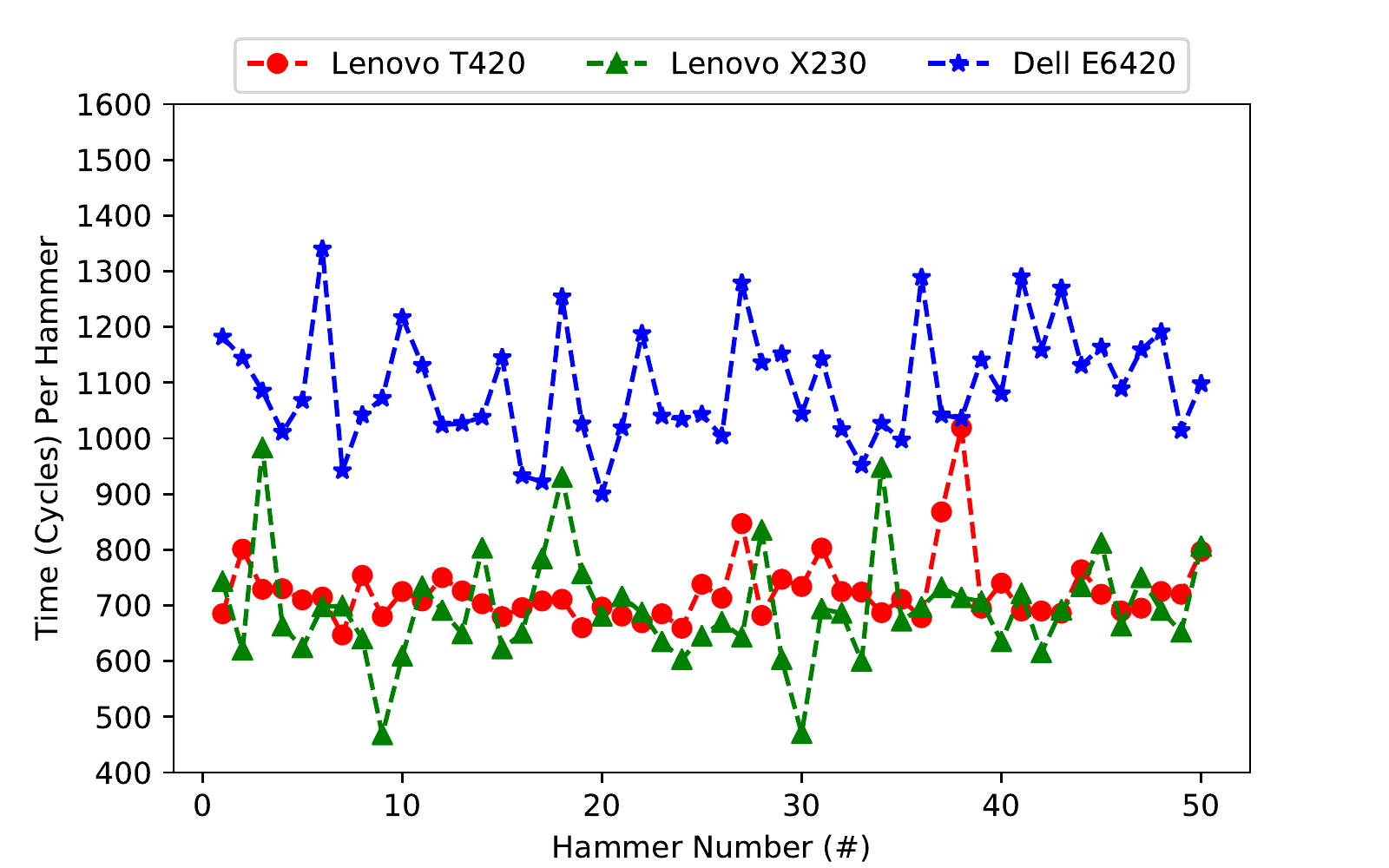}
		\caption{Double-sided hammering with superpages.}
		\label{fig:2MBcyclerange}
	\end{subfigure}%
	\caption{In both system settings, the time-cost range on each machine is well below the maximum time cost (see Figure~\ref{fig:distribution}) that allows bit flips.}
	\label{fig:cyclerange}
\end{figure*}
\subsection{Double-sided Hammering}\label{sec:double-sided pthammer}
To efficiently induce bit flips, we should hammer two L1PTEs that are one row
apart within the same bank, similar to the way how double-sided hammering
works.  As such, we expect to select appropriate user virtual addresses such
that their relevant L1PTEs meet the above requirement.  However, the physical
address of each L1PTE is required to know its location (i.e., DIMM, rank, bank,
and row) in DRAM given that a physical address to a DRAM location has been
reverse-engineering~\cite{xiao2016one,pessl2016drama}.  As we have no
permission to access the kernel space, we cannot know the physical address
of an L1PTE, making it challenging to conduct double-sided hammering.

\newcommand{\RowsSize}{\mathit{RowsSize}}
To address this problem, we are inspired by previous
works~\cite{seaborn2015exploiting,
cheng2018still,kwong2020rambleed,van2016drammer} and \new{follow  a two-steps approach}.
\new{In the first step}, we select a pair of addresses \new{whose} respective L1PTEs are
highly likely to be one row apart.  As the DRAM row size per row index, denoted by
$\RowsSize$, is \new{known (being 256\,KiB  on our test
machines)}, we \new{manipulate} the buddy allocator 
to \new{allocate} a large enough \new{number} of
Level-1 page tables. (We use the \texttt{mmap} system call to allocate 2\,GiB of
Level-1 page-tables out of the total 8\,GiB DRAM, 8\,K times as large as
$\RowsSize$.)
\new{Because the Linux buddy allocator tends to allocate consecutive physical memory pages,
many of the allocated L1PTs are in consecutive pages.
We now choose two virtual addresses that differ by
$2 \cdot \RowsSize \cdot 512$ bytes, or 256\,MiB on our test machines.
Each level-1 page table (L1PT) page contains 512 entries, 
each mapping 4\,KiB of virtual addresses.
Hence, assuming mostly consecutive page-table allocation,
the L1PTEs of the addresses we select are highly likely to be one DRAM 
rows apart.}

\new{In the second step we verify that the L1PTE pairs we found in the first
step are in the same bank.
For this, we rely on the timing difference between accessing memory
locations that are in the same bank vs.\ memory locations in different 
banks.
Specifically, accessing memory locations in different rows of the same
memory bank triggers a row-buffer conflict~\cite{moscibroda2007memory},
which slows accesses down.
Thus, for each address pair, we repeatedly perform TLB and LLC 
flushes to evict their L1PTEs from the TLB and the cache.
We then measure the access latency to the addresses in the pair.
If the L1PTEs are in the same DRAM bank, resolving the physical addresses
will be slower than if the L1PTEs are in different banks.

Experimentally evaluating the performance of this method, we find that over 95\% of
the address pairs that show slow access are in the same bank.
Furthermore, we find that of these address pairs whose L1PTEs are
in the same bank, 90\% are indeed one row apart.}

\subsection{\upperpt Performance}\label{sec:timecosts}
As in Section~\ref{sec:pthammer}, the time cost per double-sided hammering must
be no greater than the maximum latency allowed to induce bit flips. We firstly
identify the maximum time cost that permits a bit flip on each machine through
a published double-sided hammering tool\footnote{
  \url{https://github.com/google/rowhammer-test}}. 

The tool embeds two \texttt{clflush} instructions inside each round of
double-sided hammering.  To increase the time cost for each round of hammering,
we add a certain number of \texttt{NOP} instructions preceding the
\texttt{clflush} instructions in each run of the tool. We incrementally add the
\texttt{NOP} number so that the time cost per hammering will grow. The time
cost for the first bit flip to occur on each machine is shown in
Figure~\ref{fig:distribution}. As shown in the Figure, the time cost until the
first bit-flip increases with an increasing cost per hammering.  When the time
cost per hammering is more than 1500 cycles on both Lenovo machines while 1600
on the Dell machine, not a single bit flip is observed within 2 hours. As such,
1500 and 1600 are the maximum cost permitted to flip bits for the Lenovo and
Dell machines, respectively. 

We then check whether the time taken by each round of double-sided hammering is
no higher than the permitted cost. For each double-sided hammering, it
requires accesses to two user virtual addresses as well as their respective TLB
eviction set (i.e., 24 virtual addresses in total on each machine) and LLC
eviction set (i.e., 26 virtual addresses on each Lenovo machine and 34 virtual
addresses on the Dell machine).  In each system setting, we conduct
double-sided hammering for 50 rounds on each machine and measure the time that
each hammering takes.  As Figure~\ref{fig:4kbcyclerange} shows, \new{the vast majority
of double-sided hammering attempts in both Lenovo machines are in
the range of 600--900 cycles. (100\% are below 1000 cycles.) For the
Dell machine, the range is 900--1400 cycles.
When using superpages (Figure~}\ref{fig:2MBcyclerange}\new{),
94\% of the double-sided hammering attempts take 400--900 cycles  in both Lenovo
machines with an upper bound of 1100 cycles.
On the Dell machine, the range is 900--1400 cycles.}  Clearly,
the time taken per double-sided hammering is well below the maximum cost in
Figure~\ref{fig:distribution}, making \upperpt fast enough to induce bit flips.
Also, the low time cost implies that most address accesses within each
hammering are served by CPU caches rather than DRAM.

\subsection{Kernel Privilege Escalation}\label{sec:kernel_privilege}
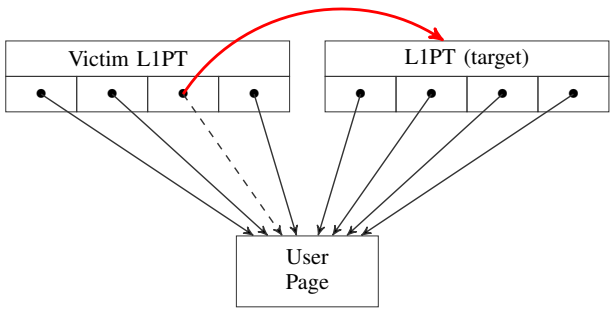
\begin{figure}[htb]
  \centering
  \resizebox{0.95\hsize}{!}{%
  \begin{tikzpicture}[thick, draw=black!80, line width=0.1mm]
    \small
    \tikzstyle{box} = [rectangle,text centered, draw=black!80, align=center]
    \path (2.25,-1) node[box, minimum width=2cm,minimum height=1cm] (UP) {User\\Page};
    \path (0,2) node[box, minimum width=4cm, minimum height=0.5cm] (PM) {Victim L1PT~~~~~};
    \path (-1.5,1.5) node[box, minimum width=1cm, minimum height=0.5cm] (PM1) {$\bullet$};
    \path (-0.5,1.5) node[box, minimum width=1cm, minimum height=0.5cm] (PM2) {$\bullet$};
    \path (0.5,1.5) node[box, minimum width=1cm, minimum height=0.5cm] (PM3) {$\bullet$};
    \path (1.5,1.5) node[box, minimum width=1cm, minimum height=0.5cm] (PM4) {$\bullet$};
    \draw[-stealth',line width=0.2mm] (PM1.center) -- (UP);
    \draw[-stealth',line width=0.2mm] (PM2.center) -- (UP);
    \draw[-stealth',dashed,line width=0.2mm] (PM3.center) -- (UP);
    \draw[-stealth',line width=0.2mm] (PM4.center) -- (UP);
    \path (4.5,2) node[box, minimum width=4cm, minimum height=0.5cm] (PPM) {L1PT (target)};
    \path (3,1.5) node[box, minimum width=1cm, minimum height=0.5cm] (PPM1) {$\bullet$};
    \path (4,1.5) node[box, minimum width=1cm, minimum height=0.5cm] (PPM2) {$\bullet$};
    \path (5,1.5) node[box, minimum width=1cm, minimum height=0.5cm] (PPM3) {$\bullet$};
    \path (6,1.5) node[box, minimum width=1cm, minimum height=0.5cm] (PPM4) {$\bullet$};
    \draw[-stealth',line width=0.2mm] (PPM1.center) -- (UP);
    \draw[-stealth',line width=0.2mm] (PPM2.center) -- (UP);
    \draw[-stealth',line width=0.2mm] (PPM3.center) -- (UP);
    \draw[-stealth',line width=0.2mm] (PPM4.center) -- (UP);
    \path (PM3.center) edge[-stealth',line width=0.4mm, draw=red, bend left=45]  (PPM);
  \end{tikzpicture}
  }
  \caption{\new{Kernel privilege escalation. Implicitly hammering L1PTEs flips a bit in a victim L1PTE, resulting in a user page map to a target L1PT page. The attacker can now modify the target L1PT page and achieve access to any desired physical memory page.}
  \label{fig:kernelprivesc}}
\end{figure}
\new{
Experimenting in both system settings, we find that \upperpt can cause a
bit flip within 15 minutes or less. (Average of five tests.)
We can identify bit flips in L1PTE by comparing the contents of the memory~\cite{cheng2018still}.

Specifically, we use the \texttt{mmap} system call to create a large number of
virtual addresses that all map to the same physical frame at user space. 
(See Figure~\ref{fig:kernelprivesc}.)
In practice, due to the limitation of the number of \texttt{mmap}ed regions,
we have several user pages, each mapped multiple times.
Such allocations create a large number of Level-1 Page Table (L1PT) pages.

We then use \upperpt to implicitly hammer such L1PT pages.
After hammering two L1PTEs each time, we check the contents of the pages in 
the attacker's address space.
In the case of a successful hammering, one or more of the L1PTEs in the victim L1PT page will experience a bit flip that will change the physical frame it points to.
We identify such success by comparing the contents of the pages in the address
space against the known contents of the user pages.

As depicted in Figure~\ref{fig:kernelprivesc}, 
because there are many L1PT pages in our address space,
there is some non-negligent probability that the frame the
modified L1PTE points to contains another L1PT.
We identify this case by checking for known patterns in L1PT pages,
and we verify this case by modifying the contents of the page and
checking for further changes in our address space mappings.

After we gain access to an L1PT page, we modify the part of the address space that this L1PT maps to point to any desired physical memory frame, achieving complete control of the system.
We note that CTA~\cite{wu2018CAT} protects against getting access to L1PTs.
See Section~\ref{sec:CTA} below for kernel privilege escalation with CTA.
}

\subsection{Software-only Rowhammer Defenses}\label{sec:defenses}
We now evaluate three existing software-only defenses
CATT~\cite{brasser17can}, RIP-RH~\cite{bock2019rip}, and CTA~\cite{wu2018CAT} against \upperpt.

\subsubsection{CATT}
CATT (CAn't-Touch-This)~\cite{brasser17can} aims to protect the 
kernel from rowhammer attacks.
It partitions each DRAM bank into kernel and user parts, reserving a small number
of rows as a buffer between the parts.
Because the user only has access to user pages, CATT deprives the attacker's
access to \emph{exploitable} hammer rows.  
The rows that the attacker is allowed to access are never adjacent to rows
that contain kernel data.

Because the page tables are kernel data, they are stored in the kernel part of the memory.
Thus, \upperpt can exploit the page-table walk to implicitly hammer rows in the
kernel part of the memory.
Furthermore, because the page tables, which we hammer, can only be in a restricted
part of the memory, selecting L1PTEs at random has a \emph{higher} probability of
picking a pair at the two sides of a victim row than if L1PTEs spread all over the memory,
increasing the chance of a successful double-hammer.
Furthermore, because the victim row is in the same limited space,
it has a higher probability of containing an L1 page table, 
increasing the likelihood of a successful attack.

We could not obtain CATT for evaluation.
Instead, we use a technique due to Cheng et al.~\cite{cheng2018still} for
increasing the concentration of L1PTEs in memory.
Specifically, we \new{exploit} the buddy allocator in the Linux kernel by first
exhausting all small blocks of memory and then starting to allocate L1PTEs.
We then run \upperpt, achieving privilege escalation within three bit flips.

\subsubsection{RIP-RH}
RIP-RH~\cite{bock2019rip} aims to isolate users by 
segregating their memory into dedicated areas in the DRAM, 
preventing cross-user rowhammer attacks.
As RIP-RH does not protect the kernel, our attack trivially applies
to it.
The code for RIP-RH is not available, but we
suspect that isolating user processes means that the kernel is
concentrated in a small part of the memory. 
Thus, we suspect that, like CATT, RIP-RH \emph{increases} the efficiency of
\upperpt.

\subsubsection{CTA}\label{sec:CTA}
CTA (Cell-Type-Aware)~\cite{wu2018CAT} employs a multi-layer approach for protecting L1PTEs
from rowhammer attacks.
In the first layer, similar to CATT, CTA segregates the L1PTs into a 
dedicated region of memory.
A further layer of defense ensures that even in the case of a bit flip in an L1PTE the
user will not get unfettered access to an L1PT page.
To achieve that, CTA places the L1PTEs in the higher addresses of the physical memory,
and verifies that the rows it uses for the L1PTEs only contain
\emph{true cells}~\cite{kim2014flipping}, i.e., memory cells that might 
change from 1 to 0 but not vice versa.
This property ensures that even if a bit flips the new address will be lower than the original
address.
Because the physical addresses of the L1PTs are all higher than all of the addresses 
of user pages, a bit flip cannot change an L1PTE from pointing to a user page
to pointing to an L1PT.

Clearly, \upperpt can overcome the first layer of defense in CTA.
However, the second layer presents a challenge.
To overcome this, we note that CTA only protects the L1PTEs, but does not
protect any of the other kernel pages.
We therefore suggest adopting a prior attack on user credentials~\cite{cheng2018still}.
Specifically, we create a large number of processes, ``sprinkling'' the kernel
memory with \texttt{struct cred} entries.
We then deploy \upperpt to flip a bit in an L1PTE.
As discussed, such a flip will not allow access to a page-table page.
However, with some non-negligible probability, it will give us access to a page that contains the \texttt{struct cred} of one of the processes
we created.
We can then change the credentials and achieve privilege escalation.

We could not obtain the CTA source code for evaluating our attack.
Instead, we simulate the attack on an undefended kernel.
We created 32\,000 processes and performed a \upperpt attack.
When a bit flip occurs in a PTE,
we search for \texttt{struct cred} in the page we gained access to.
We recognize these pages by looking for the known user ids and group ids
stored in the \texttt{struct cred}.
In our experiments, we gain \texttt{root} privileges after seven bit flips.

\section{Discussion}\label{sec:dis}

\eat{
\mypara{Defeat ZebRAM~\cite{konoth2018zebram}}
ZebRAM is a rowhammer defense but only works for a virtualized system. We can extend \upperpt to defeat it in our future work.

Empirically, ZebRAM  observes that hammering a $row_i$ can only affect  adjacent $row_{i+1}$ and $row_{i-1}$. Based on this observation, ZebRAM leverages the hypervisor to split memory of a VM into safe and unsafe regions using even and odd rows in a zebra pattern. That is, all even rows of the VM are for the safe region that contains data, while all odd rows are for the unsafe region as swap space. As such, a rowhammer attack from the safe region can only incur useless bit flips in the unsafe region. For a rowhammer attack from the unsafe region, it is not possible since the unsafe region is inaccessible to an unprivileged attacker.   

However, our experimental results in Lenovo X230 show that $row_{i+2}$ and $row_{i-2}$ are able to induce bit flips in $row_{i}$, to be specific, 173 bit flips have occurred in $row_{i}$ within 16 hours. 
Clearly, ZebRAM's observation does not hold at least in our test machine and thus enabling both \upperperi and \uppertele to defeat ZebRAM in such a machine. 
Also, Kim et al.~\cite{kim2014flipping} report that ZebRAM's observation is not correct, i.e., hammering a row can affect three rows or more in a certain number of DRAM modules.

For those modules that support ZebRAM's empirical observation, an attacker can compromise ZebRAM as follows. ZebRAM does not protect the physical memory of the hypervisor and thus extended page tables (EPTs) residing in the hypervisor space are adjacent to each other. As such, an unprivileged attacker can initiate regular memory accesses to conduct \upperpt-like attacks, causing exploitable bit flips in EPT entries and escaping the VM. 
}
\mypara{Limitation}
\upperpt does not overcome the ZebRAM~\cite{konoth2018zebram} defense.
ZebRAM targets virtualized environments, but has a high performance
overhead~\cite{wu2018CAT}, limiting its practicality. 
\new{Moreover, ZebRAM relies on the unproven assumption that hammering only
affects neighboring rows.
Thus, it ignores the possibility of 
DRAM row remapping~\cite{wu2018CAT}
or of flips happening further away from the hammered rows~\cite{kim2020revisiting}.}


\mypara{Other Possible Instances of \UPPERTELE}
Besides \upperpt, there might also exist other instances of \uppertele that leverage other built-in features of modern hardware/software. 
Particularly, the features that focus on functionality and performance may become potential candidates.  
For the hardware, we discuss two popular CPU features. Specifically, out-of-order and speculative execution are two optimization features that allow parallel execution of multiple instructions to make use of instruction cycles efficient. 
As such, an unprivileged attacker can leverage such hardware features to bypass the user-kernel privilege boundary and access kernel memory ~\cite{kocher2018spectre,lipp2018meltdown}.
Kiriansky et al.~\cite{kiriansky2018speculative} hypothesize that speculative execution might be used to mount a rowhammer attack, but they didn't have a further exploration. 

As for software, we identify some OS kernel features that may be exploitable for \uppertele. 
\eat{A system call is a programmatic interface in which a user application requests a service from the kernel.} By invoking a system call handler, a user indirectly accesses the kernel memory. Konoth et al.~\cite{konoth2018zebram} attempted performing a syscall-based rowhammer attack but didn't succeed even in an experimental attack scenario (i.e., with kernel privilege to flush target addresses) because that their hammering was inefficient.
A network I/O mechanism is also a programmatic OS feature that serves requests from the network. Particularly, the network interface card (NIC) throws an exception to notify the kernel of each network packet NIC receives. Within the exception handler, the kernel will access kernel memory. Thus, a remote user invokes this feature to access kernel memory. 

As a result, an attacker needs not only implicit but also frequent DRAM accesses to target addresses. 

\mypara{Compare \upperpt with CATTmew~\cite{cheng2018still}}
CATTmew can compromise last-level PTEs using kernel memory buffers. 

\newpar{
\mypara{Hardware Variations}
Modern Intel processors support non-inclusive caches.
Such caches are known to limit cache-based side channel attacks, such as
Flush+Re\-load~\cite{YaromF14} and Prime+\allowbreak Probe~\cite{liu2015last}.
Because in our attack we only evict data that belongs to us, we do not
expect this data to be in other cores. 
Hence, evicting it from the LLC will force future memory accesses
even when the LLC is non-inclusive.
Moreover, these non-inclusive caches are vulnerable to directory attacks~\cite{YanSGFCT19},
which we could use for \upperpt.

As Section~\ref{sec:timecosts} discusses, \upperpt can be slowed down substantially,
while still allowing hammering.
Consequently, we believe that \upperpt will be effective in machines that
require somewhat longer time for eviction, e.g., due to associative TLB
or larger associativity in the LLC.
Cache designs that aim to prevent the creation of eviction sets,
such as CEASER~\cite{Qureshi18} or ScatterCache~\cite{WernerUG0GM19},
or that randomize the TLB~\cite{deng2019secure}
can prevent \upperpt.
To the best of our knowledge, no mainstream processor implements 
such an approach.

We have only experimented with DDR3.
Recent works~\cite{kim2020revisiting,frigo_trrespass_2020} show that DDR4 is more vulnerable than DDR3,
significantly reducing the required number of accesses to the aggressor rows.
Frigo et al.~\cite{frigo_trrespass_2020} further show how to bypass the Target Row Refresh (TRR) 
rowhammer defense.
Consequently, we believe that \upperpt is applicable to DDR4 memory as well.}

\mypara{Mitigations}
Existing hardware schemes against the rowhammer attacks have two categories,
counter-based Row
Activation~\cite{kim2014architectural,seyedzadeh2016counter,seyedzadeh2018mitigating,DDR4,
LPddr4,lee2019twice} and probabilistic protection
solutions~\cite{kim2014flipping,son2017making}. The first records the number of
\texttt{ACT} commands that are sent to the same rows and starts refreshing
adjacent rows when the \texttt{ACT} number exceeds a threshold. The other
category probabilistically refreshes adjacent rows of activated rows.  However,
they require new hardware designs and thus cannot be backported to legacy
systems. 
\new{Also note that Target Row Refresh (TRR), a common
counter-based rowhammer countermeasure has been shown to be
insufficient~\cite{frigo_trrespass_2020}.
Thus it is not clear whether counter-based solutions can prevent rowhammer attacks.}

Also, there are detection-based software approaches. One is performance-counter
based such as Anvil~\cite{aweke2016anvil} and in-kernel Rowhammer
defense~\cite{kernelhammer}. They monitor the cache miss rate to detect an
ongoing rowhammer attack.
\new{We note that Anvil compares the load addresses to detect same-row accesses,
and will have to be extended to also check the L1PTE addresses to detect \upperpt.}
The other approach is RADAR~\cite{zhangleveraging} that leverages specific
electromagnetic signals. RADAR observes that rowhammer attacks exhibit
recognizable hammering-correlated sideband patterns in the spectrum of the DRAM
clock signal.  Both detection approaches require actions for preventing
hammering whenever suspicious DRAM accesses occur, resulting in substantial
performance overhead~\cite{lee2019twice}.

Besides, \upperpt is a kind of eviction-based cache and TLB attacks. It
\new{exploits} the fact that cache and TLB are shared between sensitive data and
crafted user data, clearly existing cache and TLB defenses such as
Catalyst~\cite{liu2016catalyst}, ScatterCache~\cite{werner2019scattercache} and
Secure TLBs~\cite{deng2019secure} can also mitigate \upperpt by partitioning or
randomizing either cache or TLB. 

\new{Finally, as \upperpt relies on memory spraying, monitoring the virtual address space
can detect \upperpt.  
It may be possible to use \upperpt with a smaller memory signature, at an increased
complexity for finding an exploitable bit flip. 
This increases the risk of detection through monitoring for sporadic system errors.}

\section{Conclusion}\label{sec:conclusion}
In this paper, we first observed a critical condition required by existing
rowhammer exploits to gain the privilege escalation or steal the private data.
We then proposed a new class of rowhammer attacks, called \uppertele, that
crosses privilege boundary and thus eschews the condition. 

On top of that, we created a practical instance of \uppertele, called \upperpt
that could cross the user-kernel boundary and induce an exploitable bit flip in
one Level-1 page table entry to gain kernel privilege. The experimental results
on three test machines showed that the first cross-boundary bit flip occurred
within 15 minutes of double-sided hammering. We also evaluated three DRAM-aware
software-only defenses against \upperpt and showed that \new{it can bypass them.}

\section*{Acknowledgment}
This project was supported by an Australian Research Council  Discovery Early Career Researcher Award (project number: DE200101577) and by a gift from Intel.

\bibliographystyle{plain}
\bibliography{defs,main}

\end{document}